# HIGHER ORDER STATISTICS FROM THE APM GALAXY SURVEY

István Szapudi[1], Gavin Dalton[2]

George Efstathiou[2] and Alexander S. Szalay[1,3]

## ABSTRACT

We apply a new statistics, the factorial moment correlators, to density maps obtained from the APM survey. The resulting correlators are all proportional to the two point correlation function, substantially amplified, with an amplification nearly exponential with the total rank of the correlators. This confirms the validity of the hierarchical clustering assumption on the dynamic range examined, corresponding to $0.5\,\mathrm{h}^{-1}\mathrm{Mpc} - 50\,\mathrm{h}^{-1}\mathrm{Mpc}$ in three dimensional space. The Kirkwood superposition with loop terms is strongly rejected. The structure coefficients of the hierarchy are also fitted. The high quality of the APM catalog enabled us to disentangle the various contributions from the power spectrum, small scale nonlinear clustering, and combinatorial effects, all of which affect the amplification of the correlators. These effects should appear in correlations of clusters in a similar fashion.

[1] Dept. of Physics and Astronomy, The Johns Hopkins University, Baltimore, MD 21218

[2] Dept. of Physics, Astrophysics, University of Oxford, Nuclear & Astrophysics Laboratory, Keble Road, Oxford, OX1 3RH

[3] Dept. of Physics, Eötvös University, Puskin U. 5-7. Budapest, Hungary

# 1. INTRODUCTION

Correlation functions proved to be very useful statistics for studying the distribution of galaxies because of their relatively simple connection to both theory and observations. The two-point correlation function gives only a rather limited view of the statistics, since it is equvalent to the power spectrum with all phase information lost. If the underlying distribution is non-Gaussian, the phases are correlated, and we need higher order correlation functions for the full description. In the case of galaxy catalogs the three and four point functions are quite strong (White 1979) displaying a clear sign of non-Gaussianity. Direct measurements of the higher order correlation functions have several difficulties: these quantities depend on an increasing number of variables (of which only a few are eliminated by symmetries), and are burdened with Poisson noise, biases, subtle edge and finite size effects, etc. Most of the high quality catalogs suitable for the determination of the higher order properties of the distribution are two-dimensional projections, and deprojection is only feasible under certain simplifying assumptions, while the redshift catalogs suffer distortions caused by peculiar motions. These are among the reasons why direct determination was only successful for the three-point funtion (Peebles & Groth 1975; Groth & Peebles 1977), and for the four-point function (Fry & Peebles 1978, Sharp, Bonometto & Lucchin 1984). Efforts were therefore refocused to indirect methods, of which the ones using moment correlators and/or moments were especially successful. Szapudi, Szalay & Boschán (1992, hereafter SSB) , and Meiksin, Szapudi & Szalay (1992) used both moments and moment correlators to determine the higher order properties of the angular Lick and IRAS catalogs. Gaztañaga (1992), and Bouchet et al. (1993) used moments in cells to analyse the CfA and SSRS redshift surveys, and the IRAS galaxies with redshifts, respectively. All these references confirm the validity of a scaling law, the hierarchical assumption, defined in § 2.2. The three- and four-point functions from the Perseus-Pisces redshift survey, however, seem to support the inclusion of loop terms for a better fit (Bonometto et al. 1993). The moment methods generally determine an integral of the $N$-point correlation functions over a cell (or a pair of cells, in the case of the moment correlators), and the results can be conveniently expressed in terms of the structure constants of the hierarchical ansatz, the $Q_N$'s. The hierarchical ansatz was further supported by theoretical studies of the BBKGY equations (Davis and Peebles 1977, Fry 1984, 1986, Hamilton 1988), while it was also proposed on phenomenological grounds (Balian & Schaeffer 1989). The empirical constraints on the possible break-down of the hierarchical ansatz are, however, quite uncertain, since redshift distortions in three dimensional surveys seem to show a similar trend (Lahav et al. 1993, Vogeley et al. 1994, Matsubara & Suto 1994), although other studies suggest that the effect is negligible within the measurment errors (Fry and Gaztañaga 1994). Projection could have a similar diluting effect, but according to our simulations, it is not the case, the Kirkwood assumption with loop terms could be easily distinguished from the tree hierarchy. The subject of the present study, the APM catalog, was also analysed by Gaztañaga (1994) using moments in cells. He finds an excellent agreement with the hierarchy, and fits structure constants up to $q_9$ in a rather narrow range of angular scales $0.05° \leq \theta \leq 0.3°$. For larger scales the uncertainties limit the number of points available to fit. We present a new method, based on the work in SSB, and the theoretical foundation in Szapudi & Szalay (1993a, hereafter SS), which can handle larger angular scales as well as small scales because of the nature of the factorial moment correlators, defined in the next section. This method has the additional advantage of automatically eliminating direct contributions from Poisson noise. Our aim is to fit for the hierarchy on angular scales $0.1° - 5°$, up to tenth order. The regular moments are incorporated in the normalized factorial moments, $w_{n0}$, while new information is contained in the correlators $w_{nm}$, both defined in the next section. In §2. we define the statistics used, and briefly describe the method, in §3. we review the raw results obtained from the APM survey, §4. describes the fit for the hierarchy, in §5. we present the results of the fit, and the deprojection using all the maps, and in the final section we discuss and summarize the results. The Appendix contains some theoretical calculations which would have interrupted the flow of the main text, nevertheless they are needed for understanding and interpreting the results.

## 2. DESCRIPTION OF THE METHOD

### 2.1. Factorial Moment Correlators

We introduce a set of novel statistics, the factorial moment correlators, defined as

$$w_{kl}(r_{12}) = \frac{\langle (N_1)_k (N_2)_l \rangle - \langle (N)_k \rangle \langle (N)_l \rangle}{\langle N \rangle^{k+l}}, \quad k \neq 0, l \neq 0, \tag{2.1}$$

where we used the notatation $(N)_k = N(N-1)..(N-1+k)$ for the factorial moments of the counts in cells separated by a distance $r_{12}$. All these functions will go to zero when $r_{12}$ becomes much greater then the correlation length, since we subtracted the appropriate asymptotes. For a projected catalogue (as the APM survey, used throughout this paper) the distance between two cells is of course measured in angles. Similarly, we define the $k$th factorial moments of the cell counts, denoted as $w_{k0}$

$$w_{k0} = \frac{\langle (N)_k \rangle}{\langle N \rangle^k}, \tag{2.2}$$

where there is no subtraction. The statistic is a convenient refinement of the (regular) moment correlators used by SSB with the addition that the factorial moments automatically eliminate the terms due to Poisson noise from the discreteness of the galaxy counts, since continuum moments become factorial moments after the transition to the discrete picture(SS). This use of factorial moments not only significantly simplifies all the mathematical formulae, rendering practical the study of much higher moments numerically than previously was possible, but also gives the statistic a very intuitive meaning,

$$w_{kl} \approx \langle \underbrace{\rho_1 \ldots \rho_1}_{k} \underbrace{\rho_2 \ldots \rho_2}_{l} \rangle - \langle \rho_1^k \rangle \langle \rho_2^l \rangle, \qquad (2.3)$$

where $\rho$ is underlying continuum random field (galaxy density) normalized such that its average is 1. The factorial moment correlator $w_{kl}$ is a special $k+l$ order correlation function; the first cell is degenerate $k$ times, the second is $l$ times. Its two-point nature overcomes the difficulties of dealing with multidimensional quantities but the strength of the clustering signal is still preserved. Note, that the normalization follows from the continuum form.

### 2.2. Analytical Model and Projection

A useful property of the factorial moment correlators is that they can be expressed as a function of cell averages of the correlation functions up to a finite rank $k+l$. If $P(x)$ is the generating function of the probability distribution, and $P(x, y)$ is the bivariate generating function of the two-point distribution, the normalized factorial moments $w_{0k}$ are generated (exponentially) by

$$W(x) = P(\frac{x}{n} + 1), \qquad (2.4)$$

where $n = \langle N \rangle$, the average density, and the factorial moment correlators, $w_{kl}$, are generated (exponentially) by

$$W(x, y) = P(\frac{x}{n} + 1, \frac{y}{n} + 1) - P(\frac{x}{n} + 1)P(\frac{y}{n} + 1), \qquad (2.5)$$

see SS for details. This general form could be used to fit for the integrals of the correlation functions over cells, but a further simplifying assumption can be taken. The hierarchical ansatz for the higher order correlation functions is defined as

$$\xi^{(N)} = Q_N \sum_{trees} \xi(r_1) \ldots \xi(r_{N-1}). \qquad (2.6)$$

This is a Meyer expansion with tree level terms only. The symbolic summation means the following: we can imagine a tree graph in space using the $N$ arguments of $\xi^{(N)}$ as vertices. For each edge connecting the points $r_i, r_j$ we multiply the $N-1$ two-point correlation functions $\xi^{(2)}(r_{ij})$. Finally we sum over all possible trees. Here only the functional form is fixed, and the $Q_N$ are structure constants. Note, that this assumption can be generalized to have different $Q_N$'s associated with different graph types, or one can assume only the scaling described by the equation (Balian and Schaeffer 1989). By using the form of eq. (2.6) we essentially calculate combinatorially weighted averages for each $N$. In the next section we will show, that the hierarchical assumption is valid for the APM data in the dynamical range of our study.

Under this ansatz the generating function takes the following form (SS):

$$P(x) = \exp \sum_{N=1}^{\infty} \Gamma_N (x-1)^N Q_N, \tag{2.7}$$

where we introduced the following shorthand notation (SS)

$$\Gamma_N = \frac{N^{N-2} \xi_s^{(N-1)} (nv)^N}{N!}, \tag{2.8}$$

where $\xi_s$ is the average of the two-point correlation in a cell (the same as $\bar{\xi}$ for the one point distribution), and $v$ is the volume considered. If we denote the ('long') two-point correlation function between the two cells as $\xi_l$

$$P(x,y) = P(x)P(y)\left[1 + R(x,y)\right], \tag{2.9}$$

$$R(x,y) \simeq \xi_l \sum_{M=1, N=1}^{\infty} (x-1)^M (y-1)^N Q_{N+M} \Gamma_M \Gamma_N NM.$$

Here $R(x,y)$ is approximate: only the terms linear in $\xi_l$ were kept, since we assume that $\xi_l \ll 1$. With this

$$W(x,y) = P(\frac{x}{n}+1)P(\frac{y}{n}+1)R(\frac{x}{n}+1, \frac{y}{n}+1). \tag{2.10}$$

Expanding the generating function, the resulting formulae agree with SSB if we set their Poisson-noise term to zero. This is a useful check since that result was obtained through an entirely different method. Since we operate with finite cells, the $Q_N$'s in equations (2.7) and (2.9) are not exactly the same as in the hierarchical assumption. They involve correction factors, because we approximated the integrals of correlation functions (Balian and Schaeffer 1989, SS). The details of such a correction were worked out in Boschán, Szapudi and Szalay (1994). However, the correction factors in two dimensions are so small, that the measurement errors dominate, therefore it is not necessary include them in the calculations and in the fitting.

Under the hierarchical assumption the higher order properties of the distribution are described by a set of $Q_N$'s. Since the APM catalog is a two dimensional projection, we need to deproject the angular coefficients $q_N$ to find the spatial $Q_N$'s. This is possible via Limber's equation and the result is (Peebles 1980, SS)

$$I_N = \int d\ln y \, (y^3 \phi(y))^N y^{-\gamma(N-1)},$$
$$q_N = Q_N I_1^{N-2} I_N / I_2^{N-1}. \qquad (2.11)$$

The APM luminosity function is the usual Schechter (1976) function, with variable $M^*$ and $\alpha$, following Maddox et al. (1990a)

$$M^* = M_0^* + M_1 z \quad ; \quad \alpha = \alpha_0 + \alpha_1 z \qquad (2.12)$$
$$M_0^* = -19.8 \quad M_1^* = 1 \quad ; \quad \alpha_0 = -1, \, \alpha_1 = -2.$$

The projection coefficients are found in Table 1. They are in good agreement with the calculations of Gaztañaga (1994), where he used a magnitude cut identical to one of ours.

# 3. DATA—RAW RESULTS

The APM Galaxy Survey has been described in detail by Maddox et al. (1990b,c). The survey is constructed from machine scans of 185 survey plates from the UK Schmidt telescope in a contiguous region which is relatively free from extinction, centered on the SGP. Plate regions containing bright stars, nearby galaxies, globular clusters or plate defects are masked off from the final data. Galaxy identification is automatically performed, giving a sample which is 90–95% complete with residual stellar contamination at the 5% level for $b_J < 20.0$, rising to 10% for $b_J = 20.5$. There is also residual contamination of the galaxy sample by merged images, which is estimated to be $\approx 5\%$ in the magnitude range $17.0 < b_J < 20.0$.

We generated maps of the galaxy distribution in equal-area projections using pixel scales of 0.47°, 0.23°, 0.12°. Maps of the galaxy distribution in half-magnitude slices from $b_J = 17.0$ to $b_J = 20.5$, as well as two further maps covering $17.0 < b_J < 20.0$ and all galaxies brighter than $b_J = 20.5$ were considered for each pixel scale. A map of masked regions and survey edges was generated for each pixel scale to remove from our analysis all partially filled cells. Due to growth of empty regions as the resolution is decreased, the total area, and total number of galaxies available for analysis varies with pixel scale as given in Table 2., summarizing the most important properties of the different maps. For each density map we calculated the factorial moment correlators. Fig. 1. displays the results for the 0.47° cell size. The other cell sizes have similar appearance. Each figure displays the eight different magnitude cuts. The graphs show striking regularities. The parallel curves are the factorial moment correlators, $w_{kl}$, each corresponding to a degenerate $k + l$-point correlation function.

There are three important features to notice on the graphs: ($i$) the parallelity of the lines, ($ii$) the approximate equality of the moments corresponding to the same $k+l$, regardless of the individual values of $k$ and $l$, and, (*iii*) the steady rise of the curves: i.e. there seem to be a linear amplification in *log*-space. This latter feature is even more apparent on Fig. 2. where only the (averaged) amplitudes are shown.

### 3.1. Parallelity of the correlators

The hierarchical assumption predicts parallel curves: there must be a long correlation function dominating the behaviour (see Eq. 2.9) Conversely, if the hierarchical assumption did not hold, loop terms would appear in the Meyer expansion of the higher order correlation functions, i.e. the approximation used to calculate $R(x, y)$ would no longer be valid. The correlators could have terms with $\xi_l^2$ (or even higher order) dominating at small scales, causing an upturn. The absence of this indicates that the hierarchical ansatz is a good approximation in the dynamic range considered. The projection effects are subtle, however. Although a spatial hierarchy projects to an angular hierarchy, it is not true that any general functional form is preserved in projection. For instance let us consider the general Kirkwood approximation,

$$F^{(N)}(x_1, \ldots, x_N) = n^N \prod^{\binom{N}{2}} (1 + \xi_{ij}), \tag{3.1}$$

where $F^{(N)}$ is the disconnected correlation function. This contains loop terms, which appear as powers of $\xi_{ij}$ higher than $N - 1$. In Appendix B we calculate the factorial moment correlator under the Kirkwood superposition. It exhibits the above mentioned behaviour: for a while the curves are parallel indicating the regime for which the Kirkwood model is indistinguishable from the hierarchical assumption, and at small scales there is an upturn. This formula could be used to estimate the expected crossover scale in a spatial catalog, however, here we are dealing with an angular distribution. Although it is not possible to calculate the projection generally in this case, Tóth *et al.* (1989) found that the cubic term, characteristic of the Kirkwood assumption for the three point function, will project to a similar term divided by the sum of the angles.

Since for large scales the Kirkwood superposition converges to hierarchical clustering with $Q_N = 1$, there is a break-point where the cubic term starts to dominate the small scale behaviour, therefore we should see the upturn if our dynamic range includes the crossover scale.

In the case of the factorial moment correlators the smallest scales in consideration arise when calculating the average over a cell. It could happen that while the hierarchical approximation is good for $w_l$, it breaks down below the cell size. Then the parallelity of the curves would not be destroyed, because loop terms with $w_l^2$ would be suppressed. Other loop terms, however, could affect the integral over one cell, introducing an effective $Q_N$, changing with the cell size. This effect could manifest itself in $Q_4$ and higher terms, causing a systematic shift. There is no sign of this systematics in the APM data within the errors, meaning that we did not detect any loop turn down to the scales corresponding to the cell size.

Since an analytic formula is available only for the cubic term, it is very hard to estimate the crossover scale, although we conjecture that it would be well within our dynamic range for the higher moments, where the deviation is expected to be the most significant. The equation for a possible numerical projection can be obtained from the generating function of $w_{kl}$ as expressed in terms of the disconnected correlation functions (see Appendix B): the factorial moment correlators are essentially the integrals of the disconnected correlation functions. After the substitution of the Kirkwood superpositon and, changing the measure $dx \to d\Omega r^2 dr \psi(r) = d\mu$, where $\psi$ is the selection function, $\Omega$ is the solid angle of the pixel (see SS for details), we obtain the following equation (to be compared with Limber's equation)

$$w_{k0} = \int d\mu_1 \ldots d\mu_k \prod_{pairs} (1 + \xi_{ij}) \tag{3.2}$$

$$w_{kl} = \int_1 d\mu_1 \ldots d\mu_k \int_2 d\mu_{k+1} \ldots d\mu_{k+l} \prod_{pairs} (1 + \xi_{ij}).$$

These equations are increasingly divergent with increasing order, since e.g. for $w_{N0}$ the leading term in the integrand is $\simeq r^{-\gamma N(N-1)/2}$ while the measure is $\simeq d^{3N}r$. The correlation function, however, cannot be a true power law, there has to be a small scale cut-off. This renders the integral convergent, but its value depends sensitively on this largely unknown cut-off. If we overestimate this cut-off, we can still get a lower bound on the integral. A direct Monte-Carlo approach with the Kirkwood approximation is highly unstable, since the most important contributions to the integral arise from very unlikely configurations. Therefore we developed a projection method based on Riemannian sums, described in Appendix D. We projected $\xi(r)$, the two point function, and $w_{50}$ using both the hierarchy and the Kirkwood assumption with a certain cut-off. We have chosen $w_{50}$, because among the factorial moments this has the highest degree of divergence in the integral, and the smallest available cell size ($0.12°$), where the most prominent effect is expected. We used the $17 - 20$ magnitude slice of the catalog for these numerical studies. Fig. 3 shows the projection of the two-point function with and without a cut-off. Without smoothing, the resulting correlation function is highly consistent with the APM two-point function (Maddox *et al.* 1990c). Clearly from Fig. 3., a cut-off of $0.5\,\mathrm{h}^{-1}\mathrm{Mpc}$ is rejected by the observed APM two-point correlation function. We therefore use this value for the cut-off upper limit to produce a lower bound on the integral. The Kirkwood assumption predicts $w_{k0} \geq 1820$ (see Appendix D. for details). The observed value is $\simeq 14.7$, therefore the APM data strongly reject the Kirkwood assumption. For a hierarchy with $Q_N = 1$ we used the same code to project $w_{50}$ for a comparison. In this case there are no convergence problems, so we obtain an approximation, rather than a lower bound. The result is 7.6, while using the projection of $Q_N = 1$ and the expression in Appendix C. gives 6.9. This is fully consistent with the hierarchy, since if we perturb the $Q_N$'s slightly, we can easily obtain a value very close to the measurements. In summary, the factorial moment $w_{50}$ fully supports a hierarchy with $Q_N \simeq 1$, and strongly rejects the Kirkwood assumption.

## 3.2 Approximate Degeneracy, Amplification

If the degeneracy and the linear amplification in log space were exact we could conclude that the hierarchical structure constants satisfy $q_N = 1$ for all $N$ (see Appendix A). These properties of the correlators are, however, only approximate. Even if $Q_N = 1$ were true for the spatial coefficients, projection would modify it through eq. (2.11), although not by orders of magnitude. Since we want to consider $q_N = 1$ as a 'ground state' let us define the following quantity:

$$f_{nm} = \frac{1}{w_l}\left(\frac{w_{nm}}{nm w_{n0} w_{m0}}\right). \qquad (3.3)$$

Were the $q_N$'s all equal to one, $f_{nm} = 1$ exactly would hold through all order (see Szapudi & Szalay 1993b, hereafter SS2, or the expansion of Appendix C). Since we expect this to be only an approximation we compiled a plot of $f_{nm}$ for all of the density maps obtained from the APM survey. From Fig. 4. the deviation is less then a factor of two, so it is indeed a good approximation. According to this the correlators can be written in the following form

$$w_{nm} = n \, m \, w_l \, w_{n0} \, w_{m0} \, f_{nm}. \qquad (3.4)$$

Here the $w_l$ contains the information from the power spectrum, and that determines angular dependence of the correlators, $nm$ is a purely combinatorial factor, $w_{n0} w_{m0}$ depends on the $q_N$'s up to $\max(m,n)$ and the $w_s$, and contains the vast majority of the small scale nonlinear clustering. The corrective factor $f_{nm}$ is the residual accounting for fine details of the small scale clustering and is found to be of order unity. This is a remarkably good approximation if we consider that the amplification of the correlators ranges $3-4$ orders of magnitude, and eq. (3.4) explains it up to factor of 2. This shows that the amplification process is very robust, it depends on the details of the clustering hierarchy only subtly. These considerations are valid for the amplification of the cluster correlation function, since $w_{nn}/w_{n0}^2$ and $w_{n1}/w_{n0}$ correspond to the cluster-cluster and cluster-galaxy correlation functions in a catalog where cluster selection is modeled by a soft clipping, i.e the catalog is raised to a power (SSB). This is demonstrated on Fig. 5, where the different contributions to both the $w_{k0}$'s and $w_{kl}$'s are plotted: $(N)_k P_N/\langle N \rangle^k$. The curves are increasingly shifted toward higher values of $N$ indicating, that the most important contribution to the different moments comes from cells with higher and higher counts. Therefore raising the catalog to a power (or better replacing with a factorial moment) indeed corresponds to a contrast enhancement similar to biasing (Kaiser 1984, Bardeen *et al.* 1986). Eq. (3.4) predicts that the shape of the cluster correlation function is identical to the galaxy correlation function and the amplification is roughly $n^2$ and $n$ respectively, for the auto and cross correlation function, and that the amplification depends on the details of the selection algorithm, increasing with richness. These results are very robust since $f_{nm}$ is found to be of order unity, therefore calculations assuming $q_N = 1$ (SS2) yield very realistic results. An approximate degeneracy and linear amplification in log space can arise simply as a combinatorial effect: if we plot $k!l!$ logarithmically we obtain similar graphs. ($\log(k!l!) \simeq k(\log k - 1) + l(\log l - 1) + \log k/2 + \log l/2 + 2\log\sqrt{2\pi} \simeq (k+l)(1 + \mathcal{O}(\log l + \log k)$, via the Stirling formula). This means that $w_{k0} \simeq (k-1)!$ according to eq. (3.4), which is a good approximation from Fig. 2.

One should be aware, that the 'cluster correlation' as derived in this paper from the normalization of the factorial correlators is only an approximation to correlation functions based on more realistic cluster finding algorithms (e.g. see Dalton *et al.* 1992). Here we use distinct, fixed cell sizes, use only projected counts, and discard detailed magnitude information. In our approach, a single cluster extending over several cells would be counted with multiple weight, thus a one-to-one comparison with an objective cluster finding algorithm is not trivial. On the other hand, our approach makes an analytic understanding of the phenomenon very easy, and the gross features *i)* identical shape of $\xi_{cc}$, $\xi_{cg}$, and $\xi_{gg}$, ii) $\xi_{cg} \simeq \sqrt{\xi_{gg}\,\xi_{cc}}$ should be very robust and valid for real cluster surveys as well. Our method shows, that the actual amplification has to depend on the details of the selection algorithm, like the cluster radius and threshold, and these will differ for every cluster catalog, but the above relation should persist.

To summarize considerations in this section, in the examined regime $0.1° - 5°$ the data are well described by a hierarchy with the $q_n$'s close to 1. Since only the hierarchy is preserved by projection, we performed simulations, if projections could result in a similar behaviour in an angular catalog, even if the spatial disribution is not hierarchical. The most motivated and well defined hypothesis with loop terms, the Kirkwood superposition can be rejected by these data. The observed amplification of the correlators is disentangled into factors due to the power spectrum, the small scale non-linear clustering, combinatorial effects.

## 4. FIT

To quantify the higher order properties of the APM catalog we used the raw factorial moment correlators to fit for the $q_N$ projected hierarchical structure constants. We performed a maximum likelihood fit similarly to SSB using the hierarchical model to calculate the $w_{kl}$'s for a given set of $q_N$'s. We assumed a Cauchy distribution for the errors because of its robust handling of the outlier points (Press et al 1987). We calculated all the $w_{kl}$'s and $w_{0l}$'s for a given set of $q_n$, and a set of $\xi_s$'s for the catalogs which had the same magnitude cut but different cell sizes. This was compared to the measured factorial moment correlators from the density maps, the likelihood function was calulated, and the $q_n$'s and $\xi_s$'s were adjusted by the AMOEBA algorithm (Press et al 1987). The weights for each term in the likelihood function were assigned using a particular set of $q_n$'s, and at end of the fit we checked if the resulting $q_n$'s settled reasonably close to the ones assumed for the weights. We performed simultanious fits for a given magnitude cut with all the different cell sizes. Then we ran AMOEBA several times with random initial conditions, to assess a formal error on each $q_N$. The results are presented in Table 3. The coefficients of Table 1. can be used to convert these to spatial coefficients. The lower order $q_n$ are much better determined than the higher order ones, because the lower order ones occur in more terms (see appendix C.) then the higher order ones, therefore the algorithm is more sensitive to changing those parameters.

## 5. SUMMARY AND DISCUSSIONS

We used density maps obtained from the APM galaxy catalog to calculate factorial moment correlators. The resulting graphs showed remarkable regularity over a wide range of cell sizes and magnitude cuts: all the curves are parallel to the two point correlation function and there is a uniform amplification in log space $\log w_{kl} \simeq \mathrm{const}(k + l)$. The former points toward the validity of the hierarchical assumption for the dynamic range considered, and latter is supported by both by the $q_N$'s being not far from 1 and by combinatorial conspiracy. We fitted for the structure constants $q_N$, and used the APM luminosity function to deproject them for the different magnitude cuts. Table 3. contains the results for the different magnitude cuts, the errors $\sigma_1$ are calculated formally from hundred fits from random initial conditions. The dispersion was calculated cumulatively after each fit. We found that the errorbars and the average are sufficiently settled, therefore further runs with random initial conditions would not have given more information or tighter errorbars. In an ideal case the different magnitude cuts would give the same $Q_N$'s when deprojected: this gives a chance to give a better estimation of the errors. $\Sigma_1$ is the deprojected average random error, and $\Sigma_2$ is the dispersion calculated from the different magnitude cuts. This latter error might contain several contributions: systematic errors from the possible break down of the hierarchy, different amount of contamination for the different density maps, possible errors in the luminosity function (especially at the faint end). In general any quality or error of the data that changes with the depth can influence the final fit. Although these sources of error cannot be separated now, $\Sigma_1 + \Sigma_2$ gives probably the most realistic error estimation up to now.

We conclude that we found the validity of the hierarchical assumption over a scale range of $0.1° - 5°$. For the same range we fitted the structure coefficients which are remarkably close to 1, although the higher coefficients are also consistent with zero. This result confirms the validity of the hierarchy on scales $0.5\,h^{-1}\mathrm{Mpc} - 50\,h^{-1}\mathrm{Mpc}$, while it is inconsistant with the Kirkwood superposition on the same scales.

### APPENDIX A: AN ANALYTICAL MODEL

Let us assume that the amplification is *exactly* linear and the degeneracy is perfect, and pursue the consequences. These conditions are equivalent of $w_{kl}$ in the form $\xi_l(A/n)^{k+l}$. ¿From this the generating function of $w_{kl}$ can be evaluated as

$$\frac{W(x,y)}{\xi_l} = (e^{Ax} - 1)(e^{Ay} - 1), \tag{A.1}$$

yielding a very strong constraint for the structure constants: they have to be power law,

$$Q_N = Q_3^{N-2}, \ N \geq 2. \tag{A.2}$$

This can be proved by induction using the separability of the generating function. This would mean essentially the very attractive assumption $Q_N = 1$, since $Q_1 = Q_2 = 1$.

If we relax the condition of linear amplification, but keep the degeneracy of $k+l$, the generating function becomes of the following form

$$W(x,y) = B(x+y) - B(x) - B(y) \tag{A.3}$$

where $B(x)$ is the exponential generating function of $b(k+l) = w_{kl}$. Note, that in the former case $B(x) = \xi_l(e^{Ax} - 1)$. Of course the conditions are not exact, therefore more general forms are allowed, but nevertheless this gives a very useful toy model (see SS2 for calculation of the cluster correlations within the framework of this model), and a rough idea of the order of the $Q_N$'s.

## APPENDIX B: FACTORIAL MOMENT CORRELATORS: KIRKWOOD

To calculate the factorial moment correlators we need to express the generating functions with disconnected correlation functions, since the Kirkwood superposition is natural in those terms. The path described here briefly is analgous to the methods developed in SS, where connected correlation functions were considered. The generating functional expands in terms of the reducible correlation functions $F^{(N)}$ (no exponential),

$$Z[J] = \sum_{N=0}^{\infty} \frac{(in)^N}{N!} \int dx_1 \ldots dx_N F^{(N)}(x_1, \ldots, x_N) J(x_1) \ldots J(x_N). \quad (B.1)$$

The generating function, similarly to SS can be expressed as $P(z) = \langle e^{\epsilon_R(z-1)} \rangle$, where $\epsilon_R(x) = \int \epsilon(y) W_R(x,y) dy$ is the filtered continuum field, $W_R$ is the filter function, and $\epsilon(y)$ is the underlying random field. If we take the source function $J^*(x') = W_R(x,x')(z-1)/i$, and $W_R$ is the characteristic function of the volume considered,

$$P(z) = \sum_{N=1}^{\infty} \frac{(z-1)^N}{N!} \int_v dx_1 \ldots dx_N F^{(N)}(x_1, \ldots, x_N). \quad (B.2)$$

Similarly the bivariate generating function, $P(z_1, z_2) = \langle e^{\epsilon_R(z_1-1)\epsilon_R(z_2-1)} \rangle$, can be expressed as

$$P(z_1, z_2) = \sum_{N,M} \left\{ \frac{(z_1-1)^N (z_2-1)^M}{N!M!} \int_{V_1} dx_1 \ldots dx_N \int_{V_2} dx_{N+1} \ldots dx_{N+M} F^{(N+M)}(x_1, \ldots, x_{N+M}) \right\}. \quad (B.3)$$

According to eqs. (2.4) and (2.5) the integrals in the previous two equations are $n^k w_{k0} = \langle (N)_k \rangle$ and $n^{k+l} w_{kl} = \langle (N)_k (N)_l \rangle$. The previous two equations together with eqs. (2.4) and (2.5) mean that the factorial moments and the factorial moments correlators measure the avereges of the disconnected correlation functions over a volume or two disjoint volumes respectively.

If we substitute the Kirkwood superposition of eq. (3.1) in the first integral

$$\langle (N)_k \rangle = n^k \int dx_1 \ldots dx_k \prod_{i=1}^{\binom{k}{2}} (1 + \xi_{ab}). \tag{B.4}$$

The second integral similarly yields

$$\langle (N)_k (N)_l \rangle = n^{k+l} \int_{V_1} dx_1 \ldots dx_k \int_{V_2} dx_{k+1} \ldots dx_{k+l} \prod_{i=1}^{\binom{k+l}{2}} (1 + \xi_{ab}). \tag{B.5}$$

These equations are needed when calulating the projection of the factorial moments correlators.

## APPENDIX C: FACTORIAL MOMENT CORRELATORS: HIERARCHY

Under the hierarchical assumption the $w_{kl}$'s can be expressed finitely as a function of the average of the correlation function $\xi_s$, the 'long' correlation function $\xi_l$ and the structure constants $q_n$. The $w_{n0}$'s are obtained from the expansion of the exponential generating function eq. (2.4) as the coefficient of $x^n/n!$:

$$w_{20} = 1 + \xi_s$$

$$w_{30} = 1 + 3\,\xi_s + 3\,\xi_s{}^2\, q_3$$

$$w_{40} = 1 + 6\,\xi_s + 3\,\xi_s{}^2 + 12\,\xi_s{}^2\, q_3 + 16\,\xi_s{}^3\, q_4$$

$$w_{50} = 1 + 10\,\xi_s + 15\,\xi_s{}^2 + 30\,\xi_s{}^2\, q_3 + 30\,\xi_s{}^3\, q_3 + 80\,\xi_s{}^3\, q_4 + 125\,\xi_s{}^4\, q_5.$$

The $w_{nk}$ are obtained in a similar fashion through eq. (2.5):

$$w_{21}/w_l = 2 + 2\,\xi_s\,q_3$$

$$w_{31}/w_l = 3 + 3\,\xi_s + 6\,\xi_s\,q_3 + 9\,\xi_s{}^2\,q_4$$

$$w_{41}/w_l = 4 + 12\,\xi_s + 12\,\xi_s\,q_3 + 24\,\xi_s{}^2\,q_3 + 36\,\xi_s{}^2\,q_4 + 64\,\xi_s{}^3\,q_5$$

$$w_{51}/w_l = 5 + 30\,\xi_s + 15\,\xi_s{}^2 + 20\,\xi_s\,q_3 + 120\,\xi_s{}^2\,q_3 + 60\,\xi_s{}^3\,q_3{}^2 +$$
$$90\,\xi_s{}^2\,q_4 + 170\,\xi_s{}^3\,q_4 + 320\,\xi_s{}^3\,q_5 + 625\,\xi_s{}^4\,q_6$$

$$w_{22}/w_l = 4 + 8\,\xi_s\,q_3 + 4\,\xi_s{}^2\,q_4$$

$$w_{32}/w_l = 6 + 6\,\xi_s + 18\,\xi_s\,q_3 +$$
$$6\,\xi_s{}^2\,q_3 + 30\,\xi_s{}^2\,q_4 + 18\,\xi_s{}^3\,q_5$$

$$w_{42}/w_l = 8 + 24\,\xi_s + 32\,\xi_s\,q_3 + 72\,\xi_s{}^2\,q_3 + 24\,\xi_s{}^3\,q_3{}^2 +$$
$$96\,\xi_s{}^2\,q_4 + 24\,\xi_s{}^3\,q_4 + 200\,\xi_s{}^3\,q_5 + 128\,\xi_s{}^4\,q_6$$

$$w_{52}/w_l = 10 + 60\,\xi_s + 30\,\xi_s{}^2 + 50\,\xi_s\,q_3 + 300\,\xi_s{}^2\,q_3 + 30\,\xi_s{}^3\,q_3 +$$
$$240\,\xi_s{}^3\,q_3{}^2 + 220\,\xi_s{}^2\,q_4 + 460\,\xi_s{}^3\,q_4 + 280\,\xi_s{}^4\,q_3\,q_4 +$$
$$820\,\xi_s{}^3\,q_5 + 180\,\xi_s{}^4\,q_5 + 1890\,\xi_s{}^4\,q_6 + 1250\,\xi_s{}^5\,q_7$$

$$w_{33}/w_l = 9 + 18\,\xi_s + 9\,\xi_s{}^2 + 36\,\xi_s\,q_3 + 36\,\xi_s{}^2\,q_3 + 90\,\xi_s{}^2\,q_4 +$$
$$54\,\xi_s{}^3\,q_4 + 108\,\xi_s{}^3\,q_5 + 81\,\xi_s{}^4\,q_6$$

$$w_{43}/w_l = 12 + 48\,\xi_s + 36\,\xi_s{}^2 + 60\,\xi_s\,q_3 + 180\,\xi_s{}^2\,q_3 +$$
$$72\,\xi_s{}^3\,q_3 + 72\,\xi_s{}^3\,q_3{}^2 + 216\,\xi_s{}^2\,q_4 + 288\,\xi_s{}^3\,q_4 +$$
$$108\,\xi_s{}^4\,q_3\,q_4 + 516\,\xi_s{}^3\,q_5 + 300\,\xi_s{}^4\,q_5 + 708\,\xi_s{}^4\,q_6 + 576\,\xi_s{}^5\,q_7$$

$$w_{53}/w_l = 15 + 105\,\xi_s + 135\,\xi_s{}^2 + 45\,\xi_s{}^3 + 90\,\xi_s\,q_3 + 600\,\xi_s{}^2\,q_3 +$$
$$450\,\xi_s{}^3\,q_3 + 540\,\xi_s{}^3\,q_3{}^2 + 180\,\xi_s{}^4\,q_3{}^2 + 435\,\xi_s{}^2\,q_4 +$$
$$1410\,\xi_s{}^3\,q_4 + 645\,\xi_s{}^4\,q_4 + 1380\,\xi_s{}^4\,q_3\,q_4 + 720\,\xi_s{}^5\,q_4{}^2 +$$
$$1680\,\xi_s{}^3\,q_5 + 2040\,\xi_s{}^4\,q_5 + 540\,\xi_s{}^5\,q_3\,q_5 + 4605\,\xi_s{}^4\,q_6 +$$

$$2685\,\xi_s{}^5\,q_6 + 6630\,\xi_s{}^5\,q_7 + 5625\,\xi_s{}^6\,q_8$$

$$w_{44}/w_l = 16 + 96\,\xi_s + 144\,\xi_s{}^2 + 96\,\xi_s\,q_3 + 480\,\xi_s{}^2\,q_3 + 576\,\xi_s{}^3\,q_3 +$$

$$288\,\xi_s{}^3\,q_3{}^2 + 432\,\xi_s{}^4\,q_3{}^2 + 432\,\xi_s{}^2\,q_4 + 1152\,\xi_s{}^3\,q_4 +$$

$$144\,\xi_s{}^4\,q_4 + 864\,\xi_s{}^4\,q_3\,q_4 + 1376\,\xi_s{}^3\,q_5 + 2400\,\xi_s{}^4\,q_5 +$$

$$1536\,\xi_s{}^5\,q_3\,q_5 + 2832\,\xi_s{}^4\,q_6 + 1536\,\xi_s{}^5\,q_6 + 4608\,\xi_s{}^5\,q_7 + 4096\,\xi_s{}^6\,q_8$$

$$w_{54}/w_l = 20 + 180\,\xi_s + 420\,\xi_s{}^2 + 180\,\xi_s{}^3 + 140\,\xi_s\,q_3 + 1200\,\xi_s{}^2\,q_3 +$$

$$2340\,\xi_s{}^3\,q_3 + 360\,\xi_s{}^4\,q_3 + 1200\,\xi_s{}^3\,q_3{}^2 + 2880\,\xi_s{}^4\,q_3{}^2 +$$

$$720\,\xi_s{}^5\,q_3{}^3 + 780\,\xi_s{}^2\,q_4 + 3800\,\xi_s{}^3\,q_4 + 3300\,\xi_s{}^4\,q_4 +$$

$$4920\,\xi_s{}^4\,q_3\,q_4 + 3720\,\xi_s{}^5\,q_3\,q_4 + 2880\,\xi_s{}^5\,q_4{}^2 + 3400\,\xi_s{}^3\,q_5 +$$

$$10080\,\xi_s{}^4\,q_5 + 2040\,\xi_s{}^5\,q_5 + 9840\,\xi_s{}^5\,q_3\,q_5 + 5120\,\xi_s{}^6\,q_4\,q_5 +$$

$$10860\,\xi_s{}^4\,q_6 + 18420\,\xi_s{}^5\,q_6 + 11340\,\xi_s{}^6\,q_3\,q_6 + 24780\,\xi_s{}^5\,q_7 +$$

$$13260\,\xi_s{}^6\,q_7 + 42980\,\xi_s{}^6\,q_8 + 40000\,\xi_s{}^7\,q_9$$

$$w_{55}/w_l = 25 + 300\,\xi_s + 1050\,\xi_s{}^2 + 900\,\xi_s{}^3 + 225\,\xi_s{}^4 + 200\,\xi_s\,q_3 +$$

$$2400\,\xi_s{}^2\,q_3 + 7800\,\xi_s{}^3\,q_3 + 3600\,\xi_s{}^4\,q_3 + 3000\,\xi_s{}^3\,q_3{}^2 +$$

$$14400\,\xi_s{}^4\,q_3{}^2 + 1800\,\xi_s{}^5\,q_3{}^2 + 7200\,\xi_s{}^5\,q_3{}^3 + 1300\,\xi_s{}^2\,q_4 +$$

$$9500\,\xi_s{}^3\,q_4 + 16500\,\xi_s{}^4\,q_4 + 5100\,\xi_s{}^5\,q_4 + 16400\,\xi_s{}^4\,q_3\,q_4 +$$

$$37200\,\xi_s{}^5\,q_3\,q_4 + 13200\,\xi_s{}^6\,q_3{}^2\,q_4 +$$

$$14400\,\xi_s{}^5\,q_4{}^2 + 20800\,\xi_s{}^6\,q_4{}^2 + 6800\,\xi_s{}^3\,q_5 +$$

$$33600\,\xi_s{}^4\,q_5 + 20400\,\xi_s{}^5\,q_5 + 49200\,\xi_s{}^5\,q_3\,q_5 +$$

$$10800\,\xi_s{}^6\,q_3\,q_5 + 51200\,\xi_s{}^6\,q_4\,q_5 + 27150\,\xi_s{}^4\,q_6 +$$

$$92100\,\xi_s{}^5\,q_6 + 26850\,\xi_s{}^6\,q_6 + 113400\,\xi_s{}^6\,q_3\,q_6 +$$

$$100000\,\xi_s{}^7\,q_4\,q_6 + 82600\,\xi_s{}^5\,q_7 + 132600\,\xi_s{}^6\,q_7 + 75000\,\xi_s{}^7\,q_3\,q_7 +$$

$$214900\,\xi_s{}^6\,q_8 + 112500\,\xi_s{}^7\,q_8 + 400000\,\xi_s{}^7\,q_9 + 390625\,\xi_s{}^8\,q_{10}$$

These formulae compare directly with the formulae listed in SSB without Poisson noise, $1/N = 0$. The simplification achieved by the factorial moments can be illustrated: the $w_{42}$ term contains 23 terms if one uses regular moments as opposed to 9 for the factorial moments. The higher we go, the more dramatic is the difference due to the combinatorial explosion of terms.

### APPENDIX D: PROJECTION UNDER THE KIRKWOOD ASSUMPTION

As described in the main text the integration of Eqs. (3.2) for the projection would be highly unstable with a Monte Carlo approach. We therefore used a direct Riemannian sum to approximate the integral. The integration volume is a pyramid corresponding to the solid angle $\Omega$. We have chosen the smallest cell size (0.12°) and the $17 - 20$ magnitude slice for practical investigations. We divided this 'pencil beam' into $2^N$ slices of equal length, and on each slice we approximated the correlation function with its average on the slice. In principle, of course, we could have used any value of the integrand within the cell, but we anticipated that the average might give a more stable performance. Although the method is completely general, we used it to calculate $w_{50}$ (where the effect is expected to be the most prominent), and we describe it in those terms for simplicity. The generalization should be straightforward, although tedious for higher order factorial moments, or moment correlators. In the latter case the only difference is, that the integration volume comprises of two pencil beams.

The projection of $w_{k0}$ is a five fold integral. The five variables are distributed among the cells. There are 7 distinct cases, according to how many variables fall within one cell:

$$\begin{array}{l} 5 \\ 4\ 1 \\ 3\ 2 \\ 3\ 1\ 1 \\ 2\ 2\ 1 \\ 2\ 1\ 1\ 1 \end{array}$$

1 1 1 1 1 .

These are all the possible partitions of 5 into integers. 4 1 means that four points are in one cell, and one point is in a different cell, etc. The sum of the contributions over all possible configurations with the appropriate measure calculated from the selection function will approximate the integral as the cell size approaches zero. The problem with the Monte Carlo approach lies in the fact that the most important contribution to the integral comes from the first partition 5, where all the points lie in the same cell.

The average of the correlation function over a cell was calculated by a Monte Carlo simulation, and the following formula was found to be an adequate approximation,

$$\xi_s = \frac{7.586}{\sqrt{0.36 + 2.9x + x^2}}(d/r_0)^\gamma, \tag{D.1}$$

where the pyramid slice was approximated with a rectangular parallepiped with the ratio of the short and long side $x = d/l$, and we used $\gamma = 1.7$ and $r_0 = 5\,\text{h}^{-1}\text{Mpc}$.

The following notation is needed for the approximation of the correlation function $\xi(|\,r^{(1)} - r^{(2)}\,|)$ between two different cells (the upper index in parentheses numbers the vector, while lower index defines the coordinate). Let us draw the vector $x^{(1)}$ and $x^{(2)}$ to the center of mass of the two cell. In a general calculation these could be in two different pencil beams. Then we can write

$$r = r^{(1)} - r^{(2)} = x^{(1)} - x^{(2)} + s^{(1)} - s^{(2)} \tag{D.2}$$

where $s^{(i)}$ points from the center of mass, so its average $\langle s^{(i)} \rangle = 0$. The Taylor expansion of $\xi$

$$\xi(|\,r\,|) = \xi(R) + s_i \frac{\partial \xi}{\partial s_i}\bigg|_{s=0} + \frac{1}{2} s_i s_j \frac{\partial^2 \xi}{\partial s_i \partial s_j}\bigg|_{s=0} + \ldots, \tag{D.3}$$

where $R = |\,x^{(1)} - x^{(2)}\,|$ the distance between the center of mass of the two cells, $x = x^{(1)} - x^{(2)}$, $s = s^{(1)} - s^{(2)}$, and the Einstein convention is used. After differentiation

$$\xi(|\,r\,|) = \xi(R) + \frac{s_i x_i \xi'}{R} + \frac{1}{2} s_i s_j \left[\frac{x_i x_j}{R^2}(\xi'' - \frac{1}{R}\xi') + \delta_{ij}\frac{\xi'}{R}\right]. \tag{D.4}$$

If both of the cells are in the same beam, we can calculate the average of this using $\langle s_i s_j \rangle = \delta_{ij} \langle s_i^2 \rangle$ as

$$\langle \xi \rangle = \xi(R) + \frac{1}{2} \langle s_i^2 \rangle \left[ \frac{x_i^2}{R^2} (\xi'' - \frac{1}{R} \xi') + \frac{\xi'}{R} \right]. \tag{D.5}$$

If we substitute $R\xi' = -\gamma \xi$ and $R^2 \xi'' - R\xi' = \gamma(\gamma+2)\xi$, and we use that the two cells are in the same pencil beam therefore $x = (0, 0, R)$,

$$\langle \xi \rangle = \xi(R) - \frac{1}{2} \left( \langle (s^{(1)})^2 \rangle + \langle (s^{(2)})^2 \rangle \right) \gamma \xi'(R)/R + \tag{D.6}$$
$$\frac{1}{2} \left( \langle (s_z^{(1)})^2 \rangle + \langle (s_z^{(2)})^2 \rangle \right) \gamma(\gamma+2)\xi'(R)/R^2,$$

This correction falls off as $1/R$, and is therefore important only for the few nearest neighbours. For the touching cells, however, it is not accurate enough, and we used the constraint for the total correlation of the two cells. If $x_{11}$ and $x_{22}$ denote the averages of the correlation functions in the two touching cells, $x_{12}$ is the average of the cross correlations between them, and $\xi_0$ is the average of the total correlations in the double cell, we have the following formula because of the additivity of the integral:

$$4\xi_0 = \xi_{11} + \xi_{22} + 2\xi_{12}. \tag{D.7}$$

In the present case the two cells are nearly identical, because of the small angle of the cone, therefore $\xi_{11} \simeq \xi_{22}$.

Now to calculate the integral $w_{50} = \int d\mu_1 \ldots d\mu_5 \prod_{pairs}(1 + \xi_{ij})$ we needed to calculate the correlations on the set of cells, and the discrete integration measure $p_i$ replacing $d\mu$ was simply calculated on each cell using the selection function such that $\sum p_i = 1$ hold. When implementing it as a computer program we need only as many loops as the number of integers 5 was partitioned into, e.g. we need only one loop for the partition with all points in the same cell. If the integral is finite, the approximate sum is convergent as the cell size becomes smaller. Similarly the hierarchical $w_{50}$ could be projected using a formula similar to Appendix C, except we left $q_1$, $q_2$ in the formula explicitly, although they both equal 1, they were needed to locate the appropriate orders. Then it was a simple matter to assign the correct power of the measure and the correlation function for each term. For instance $30 q_2 q_3 \xi_s^3$ was associated with $30 p_i^2 p_j^3 \xi_i \xi_j^2$, where $i$ and $j$ denote the loop variables, etc. Finally, the same code calculated the projection of the two point function and calculated $w_s$ for mainly test purposes.

The results of the integration are the following. The two-point function and the average of the correlation function over the APM cell are well recovered by the code. The projection of $w_{50}$ under the hierarchical assumption with all $Q_N = 1$ converged at cell sizes as big as $96\,h^{-1}\mathrm{Mpc}$! The integration gave 7.6, in reasonable agreement with the formula in Appendix C. combined with the projection of the $Q_N$'s from Table 2. gave 6.92. The measured value is 14.7. The factor of two difference can be accounted for by letting the $Q_N$'s differ from 1 by a small amount. For the Kirkwood assumption we can only expect convergence if we use a cut-off in the correlation function, otherwise the integral is divergent. Every approximate sum until convergence is therefore a lower bound, and if we use a larger than real cut-off we obtain an even lower bound. The cut-off from the APM correlation function has to be smaller then $0.5\,h^{-1}\mathrm{Mpc}$. Using that as a cut-off and calculating the integral with smaller and smaller cell sizes, we did not reach convergence up to $24\,h^{-1}\mathrm{Mpc}$ cell size, but the lower bound obtained is 1820, some 124 times the observed value. Therefore the data safely reject the Kirkwood assumption.

TABLE CAPTIONS

TABLE 1.— For each magnitude cut ($m_{min} - m_{max}$) and for three different cell sizes ($l$) the average number of galaxies in a cell ($N_{ave}$) is displayed. $\xi_{est}$ is the estimated amplitude of the correlation function using only second order quantities, $\xi_{fit}$ is the result of the fit up tenth order. The fact that these two determination of $\xi$ do not differ significantly supports the validity of the fit. For the sake of completeness we give the total number of objects ($N_{tot}$), and the surface area ($A$) covered by each catalog with different cell sizes.

TABLE 2.— The coefficients used for deprojection of the angular structure constants to spatial ones. For illustration one of the magnitude cuts is calculated with both $h = 1$ (the dimensionless Hubble constant) and $h = 0.5$, the rest of the calculation $h = 0.5$ is used since the results are not sensitive to this parameter.

TABLE 3.— The main numerical results of the paper are displayed. The fitted angular coefficients $q_n$ are shown for each magnitude cut, the formal error from hundred runs with random initial conditions is $\sigma_1$. The spatial $Q_N$'s are calculated using all the angular $q_N$'s and the coefficients of Table 2. $\Sigma_1$ is the formal deprojected error, $\Sigma_2$ is the dispersion from the different deprojections used when calculating the average.

# Figure Captions

Fig. 1.— The factorial moment correlators from the catalog of 0.47° cell size are shown for each magnitude cut. The magnitude cut is coded under the figures in a way that e.g. 170200 means a cut between 17.0 and 20.0. The graphs are strikingly parallel. The correlators from the other density maps with cell sizes 0.23°, and 0.12° have similar appearance.

Fig. 2.— The amplitude $\log(w_{kl}/w_{11})$ is shown as a function of $k+l$. There is a small splitting in the degenaracy in $k+l$ compared to the average trend which is nearly linear in log-space. The different cellsizes are coded as follows: square: 0.12°, triangle: 0.23°, and cross: 0.47°. The solid line corresponds to the normalized factorial moments $w_{k0}$.

Fig. 3.— The projection of the two-point correlation function is shown with a smoothing of $a = 0.5 \, h^{-1} \mathrm{Mpc}$ and without smoothing. The projected curve without smoothing is consistent with the APM two-point function, while the other curve is strongly inconsistent. Therefore the cut-off in the real correlation function is less then $0.5 \, h^{-1} \mathrm{Mpc}$. The $w_{50}$ moment, calculated from the Kirkwood assumption with $0.5 \, h^{-1} \mathrm{Mpc}$ cut-off, is orders of magnitude larger then the observed value (which is consistent with the hierarchy). The Kirkwood assumption is strongly rejected, because smaller cut-off would result in even larger $w_{50}$.

Fig. 4.— The quantity $\log(f_{nm})$ is displayed as a function of $n+m$. This plot enables us to disentangle the different effects contributing to the amplification of the correlations. Note that $f_{nm}$ is of order unity while the amplifications on Fig. 2. are of order $10^3 \div 10^6$.

Fig. 5.— $\log((N)_k P_N / \langle N \rangle^k)$ versus $N$ is plotted for $k = 0, \ldots, 5$ to show that the factorial moments correspond to a contrast enhancement. For this plot the 17.0

to 20.0 magnitude cut with 0.23 cell size was used. The most important contribution indeed shifts towards higher $N$'s as $k$ increases.

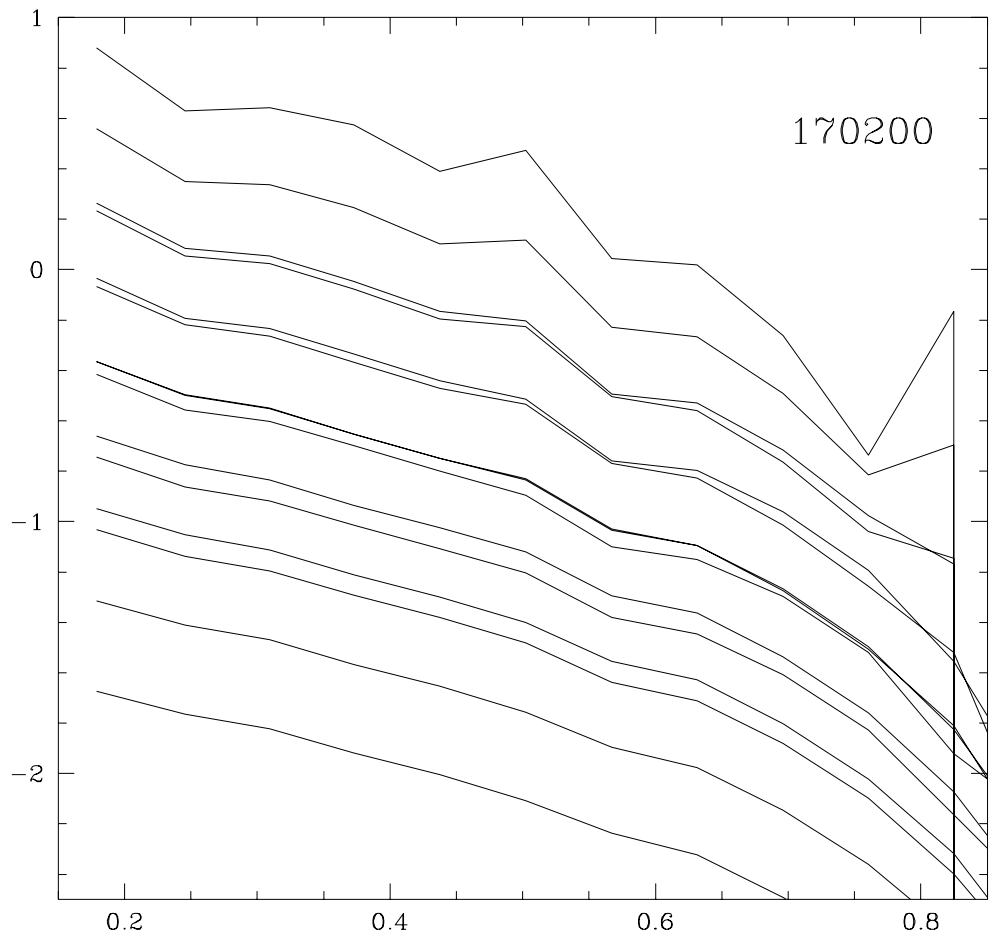
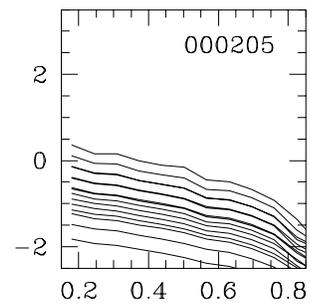
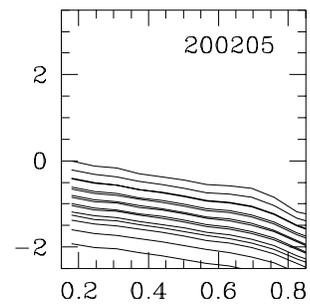
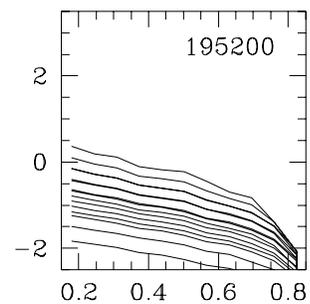
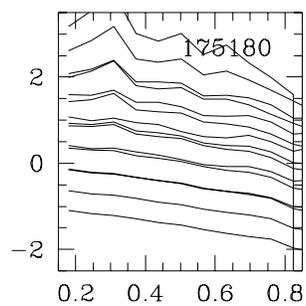
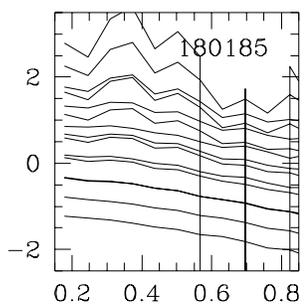
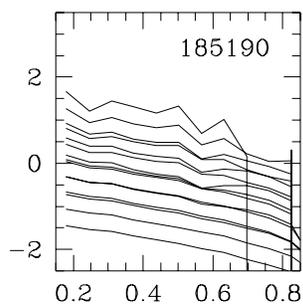
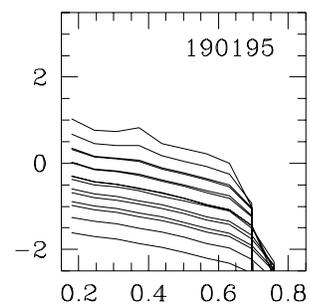

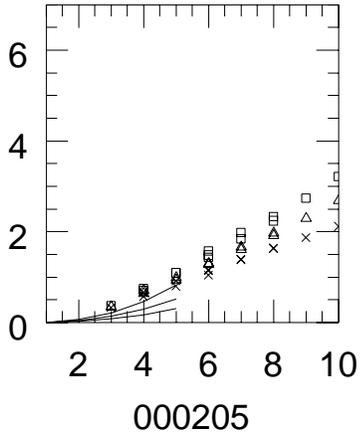

000205

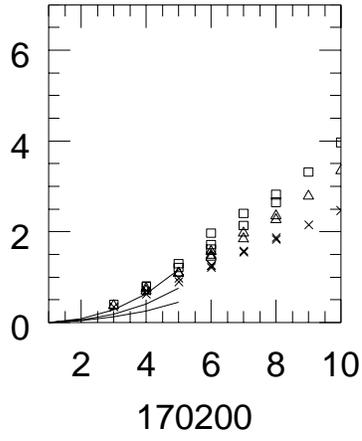

170200

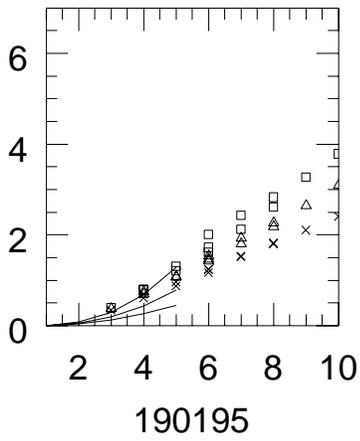

190195

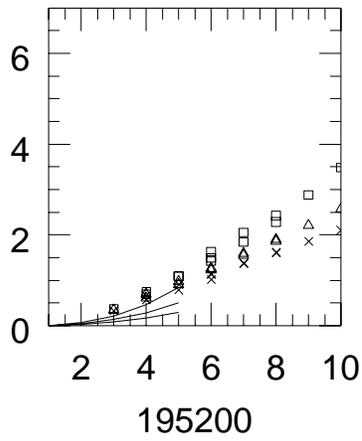

195200

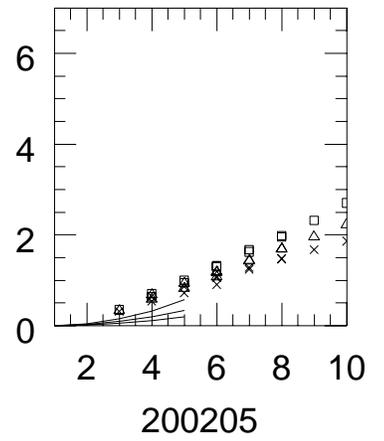

200205

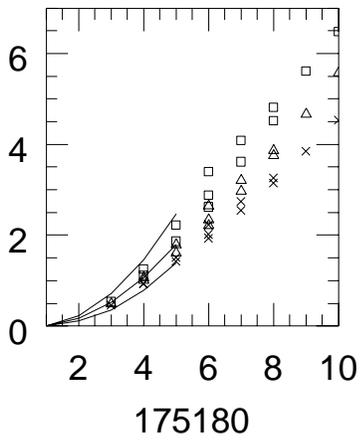

175180

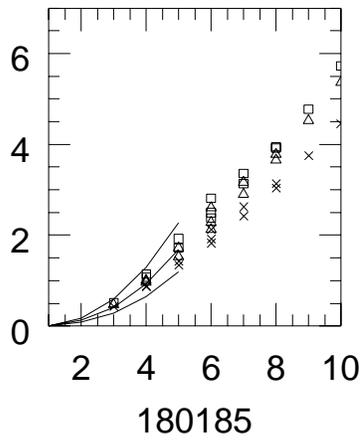

180185

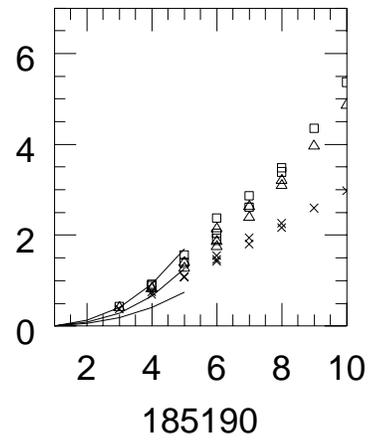

185190

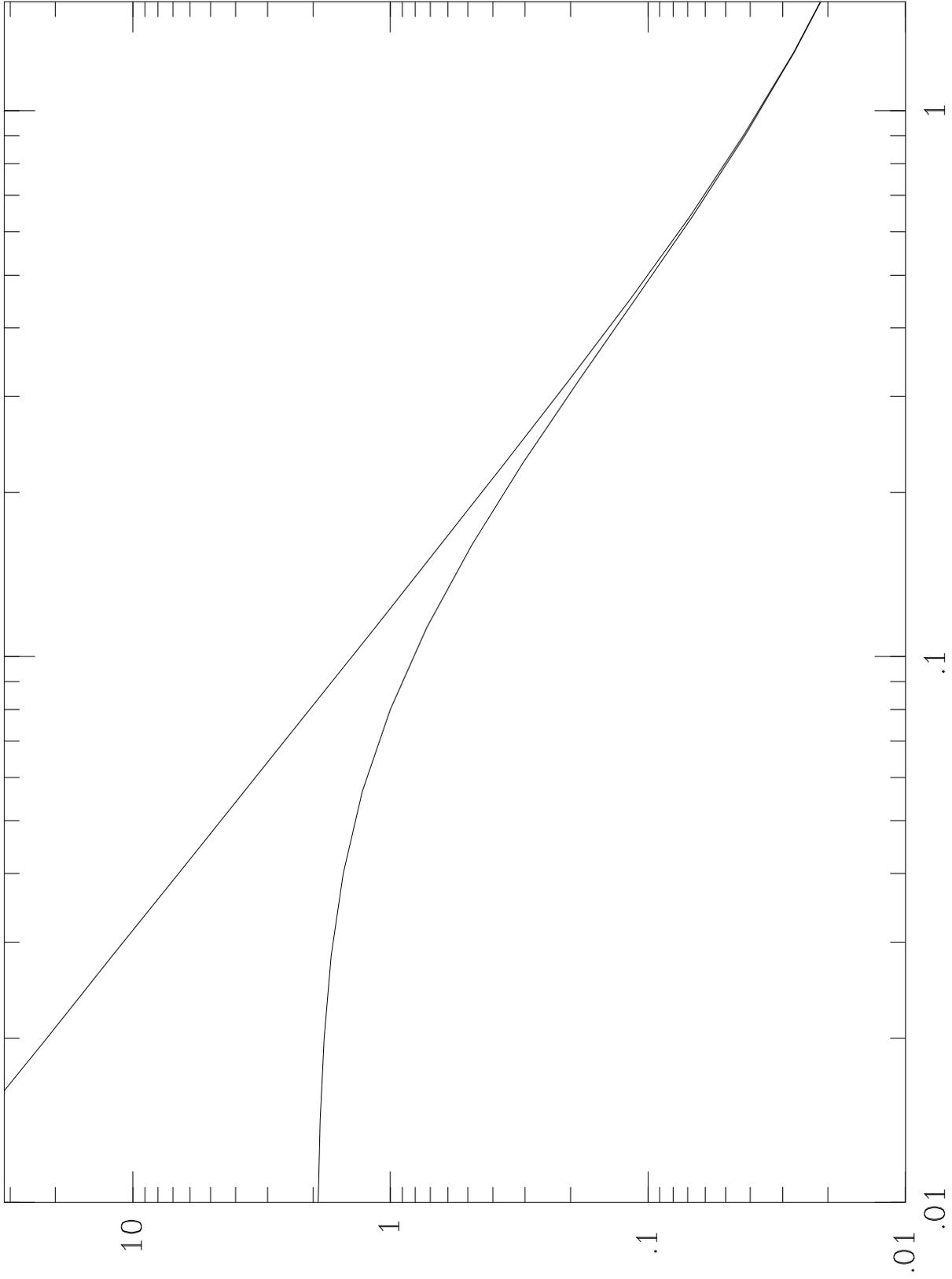

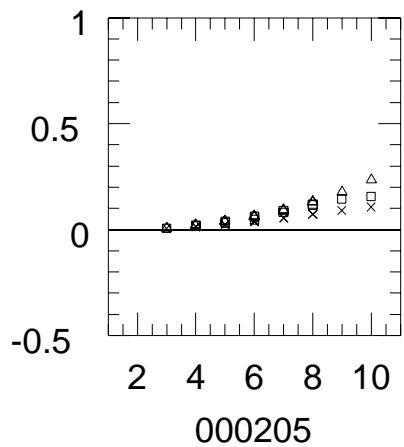
000205

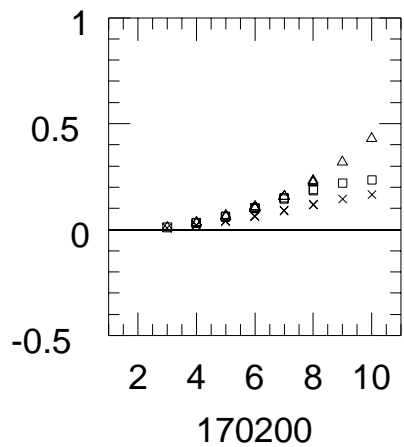
170200

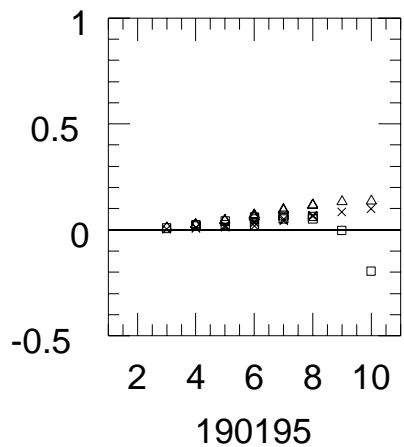
190195

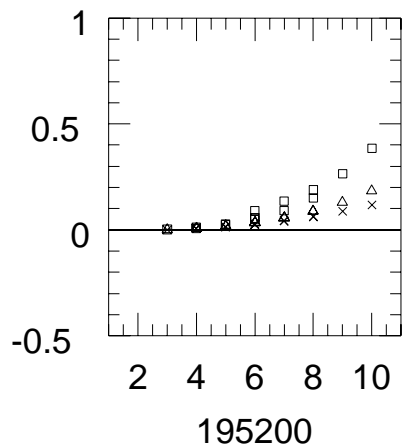
195200

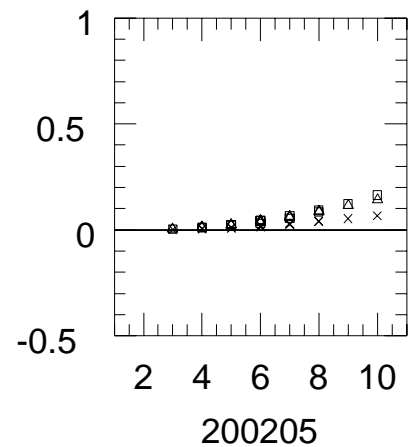
200205

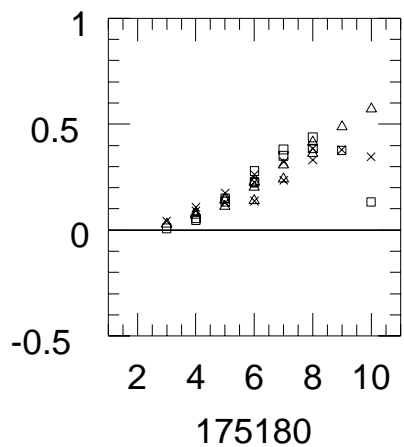
175180

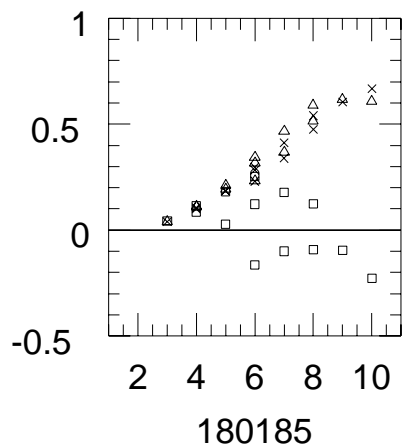
180185

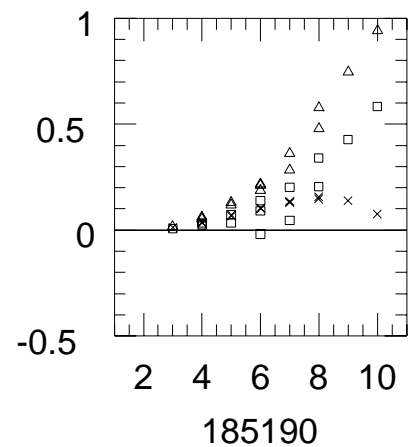
185190

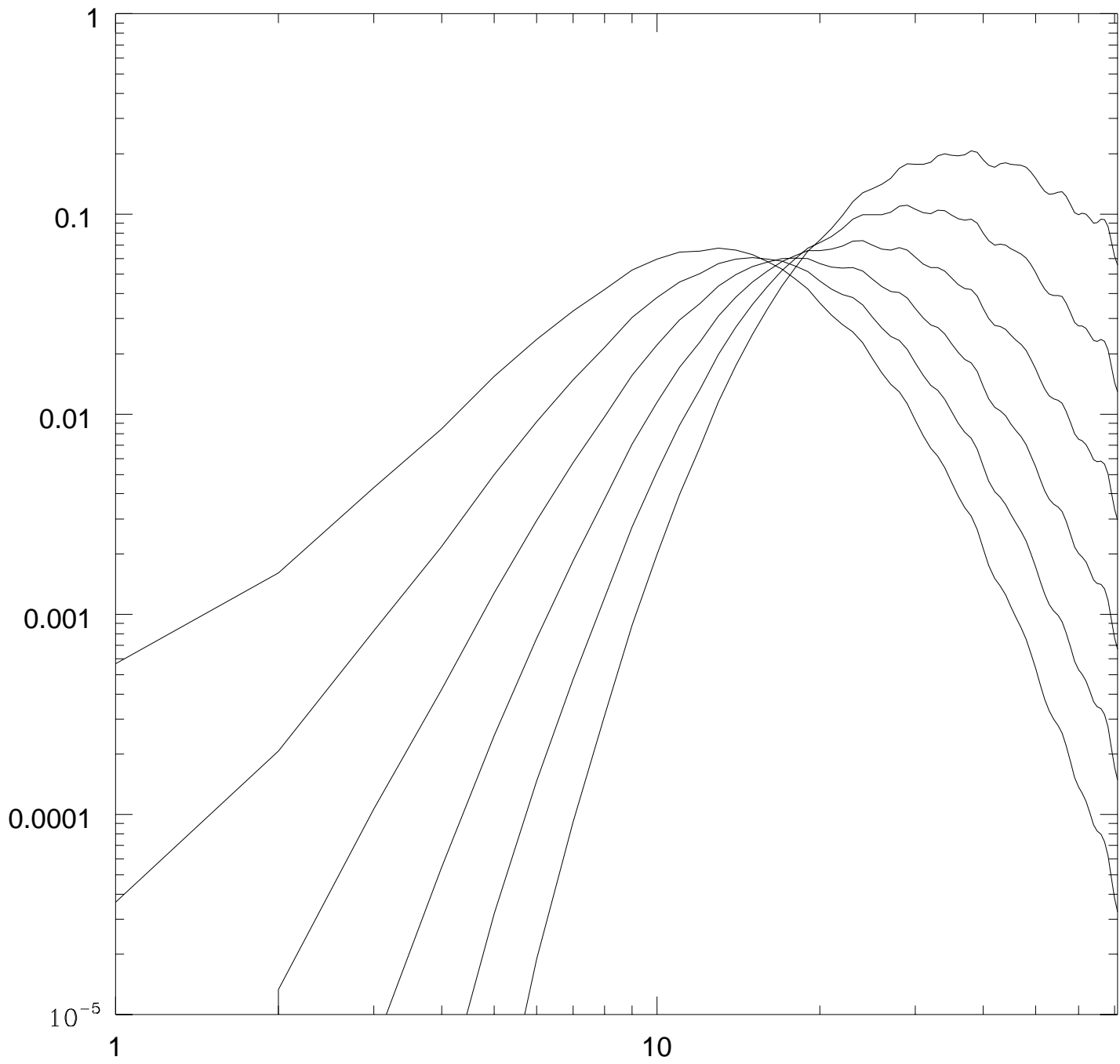

TABLE 1.

DENSITY MAPS FROM THE APM SURVEY

| $l$ | | 0.117188° | | | 0.234375° | | | 0.468750° | |
|---|---|---|---|---|---|---|---|---|---|
| $N_{tot}$ | | 2264402 | | | 2182470 | | | 1981523 | |
| $A_{tot}$ | | 4227 | | | 4069 | | | 3692 | |
| $m_{min} - m_{max}$ | $N_{ave}$ | $\xi_{est}$ | $\xi_{fit}$ | $N_{ave}$ | $\xi_{est}$ | $\xi_{fit}$ | $N_{ave}$ | $\xi_{est}$ | $\xi_{fit}$ |
| $00 - 20.5$ | 6.791 | 0.165 | 0.181 | 27.192 | 0.106 | 0.122 | 108.88 | 0.0665 | 0.0665 |
| $17.0 - 20.0$ | 3.986 | 0.224 | 0.224 | 15.961 | 0.147 | 0.158 | 63.92 | 0.0948 | 0.0949 |
| $17.5 - 18.0$ | 0.158 | 0.705 | 0.703 | 0.633 | 0.466 | 0.519 | 2.539 | 0.295 | 0.295 |
| $18.0 - 18.5$ | 0.308 | 0.504 | 0.504 | 1.235 | 0.333 | 0.424 | 4.953 | 0.224 | 0.252 |
| $18.5 - 19.0$ | 0.572 | 0.338 | 0.338 | 2.290 | 0.226 | 0.211 | 9.165 | 0.147 | 0.167 |
| $19.0 - 19.5$ | 1.072 | 0.234 | 0.229 | 4.292 | 0.154 | 0.194 | 17.20 | 0.101 | 0.105 |
| $19.5 - 20.0$ | 1.798 | 0.166 | 0.166 | 7.198 | 0.107 | 0.115 | 28.82 | 0.0678 | 0.0681 |
| $20.0 - 20.5$ | 2.713 | 0.122 | 0.122 | 10.863 | 0.0755 | 0.0757 | 43.491 | 0.0460 | 0.0459 |

TABLE 2.

CORRECTION FACTORS FOR TREE GRAPHS UP TO $N = 8$

|  | 2D | | | 3D | | |
|---|---|---|---|---|---|---|
| i/$\gamma$ | 1.5 | 1.65 | 1.8 | 1.5 | 1.65 | 1.8 |
| 1 | 1.000 | 1.000 | 1.000 | 1.000 | 1.000 | 1.000 |

TABLE 2.

PROJECTION COEFFICIENTS FOR THE $Q_N$'S

| $m_{min} - m_{max}$ | h | $R_3$ | $R_4$ | $R_5$ | $R_6$ | $R_7$ | $R_8$ | $R_9$ | $R_{10}$ |
|---|---|---|---|---|---|---|---|---|---|
| 17.0 − 20.0 | 1.0 | 1.161 | 1.434 | 1.830, | 2.386 | 3.157 | 4.221 | 5.690 | 7.719 |
| 17.0 − 20.0 | 0.5 | 1.178 | 1.483 | 1.934 | 2.579 | 3.492 | 4.781 | 6.601 | 9.174 |
| 00.0 − 20.5 | 0.5 | 1.168 | 1.453 | 1.869 | 2.454 | 3.271 | 4.405 | 5.981 | 8.171 |
| 17.5 − 18.0 | 0.5 | 1.159 | 1.426 | 1.813 | 2.351 | 3.093 | 4.110, | 5.505 | 7.420 |
| 18.0 − 18.5 | 0.5 | 1.157 | 1.419 | 1.797 | 2.323 | 3.045 | 4.033 | 5.382 | 7.228 |
| 18.5 − 19.0 | 0.5 | 1.153 | 1.411 | 1.780 | 2.293 | 2.994 | 3.950 | 5.251 | 7.025 |
| 19.0 − 19.5 | 0.5 | 1.151 | 1.402 | 1.763 | 2.261 | 2.941 | 3.863 | 5.115 | 6.813 |
| 19.5 − 20.0 | 0.5 | 1.147 | 1.393 | 1.744 | 2.228 | 2.885 | 3.775 | 4.976 | 6.600 |
| 20.0 − 20.5 | 0.5 | 1.144 | 1.384 | 1.726 | 2.196 | 2.831 | 3.688 | 4.840 | 6.391 |
| 17.0 − 17.5 | 0.5 | 1.162 | 1.433 | 1.827 | 2.377 | 3.136 | 4.181 | 5.618 | 7.596 |

TABLE 3.

Results of the Fits for the $q_n$'s

| $m_{min} - m_{max}$ | order: | 3 | 4 | 5 | 6 | 7 | 8 | 9 | 10 |
|---|---|---|---|---|---|---|---|---|---|
| 17.0 − 20.0 | $q_N$ | 1.160 | 1.964 | 5.274 | 9.900 | 2.123 | 1.236 | −0.653 | 0.380 |
|  | $\sigma_1$ | 0.002 | 0.090 | 0.625 | 2.459 | 9.208 | 9.053 | 9.728 | 11.452 |
| 00.0 − 20.5 | $q_N$ | 1.141 | 2.224 | 6.812 | 2.963 | 1.447 | 0.971 | 0.066 | 0.868 |
|  | $\sigma_1$ | 0.005 | 0.118 | 1.110 | 4.363 | 5.254 | 5.524 | 5.486 | 5.479 |
| 17.5 − 18.0 | $q_N$ | 1.624 | 3.056 | 5.868 | 12.192 | 6.035 | 3.118 | −0.197 | 3.847 |
|  | $\sigma_1$ | 0.002 | 0.009 | 0.124 | 1.215 | 14.664 | 17.136 | 21.994 | 20.197 |
| 18.0 − 18.5 | $q_N$ | 1.373 | 2.907 | 7.232 | 8.837 | 3.476 | −0.235 | 0.802 | −0.221 |
|  | $\sigma_1$ | 0.001 | 0.010 | 0.191 | 1.786 | 9.852 | 13.245 | 9.652 | 10.566 |
| 18.5 − 19.0 | $q_N$ | 1.172 | 2.466 | 5.727 | 3.312 | 1.524 | 1.445 | 0.919 | 1.313 |
|  | $\sigma_1$ | 0.002 | 0.030 | 0.303 | 1.617 | 4.922 | 5.505 | 5.010 | 5.045 |
| 19.0 − 19.5 | $q_N$ | 1.061 | 1.547 | 4.182 | 6.443 | 0.796 | −0.019 | 1.628 | 0.578 |
|  | $\sigma_1$ | 0.004 | 0.048 | 0.312 | 1.922 | 6.374 | 8.458 | 7.330 | 7.020 |
| 19.5 − 20.0 | $q_N$ | 0.939 | 1.667 | 4.892 | 5.341 | 2.004 | 1.332 | −0.124 | 0.557 |
|  | $\sigma_1$ | 0.005 | 0.125 | 1.162 | 4.906 | 8.455 | 6.342 | 6.671 | 6.514 |
| 20.0 − 20.5 | $q_N$ | 1.103 | 1.815 | 4.821 | 1.963 | 0.413 | 0.279 | 0.488 | 0.567 |
|  |  | 0.007 | 0.174 | 1.108 | 3.575 | 5.015 | 4.066 | 3.861 | 3.404 |
| total | $Q_N$ | 1.03 | 1.55 | 3.10 | 2.70 | 0.72 | 0.25 | 0.07 | 0.14 |
|  | $\Sigma_1$ | 0.00 | 0.05 | 0.34 | 1.18 | 2.59 | 2.12 | 1.59 | 1.17 |
|  | $\Sigma_2$ | 0.18 | 0.39 | 0.54 | 1.51 | 0.58 | 0.26 | 0.14 | 0.17 |